\documentclass[12pt,preprint]{aastex}
\usepackage{graphicx}
\usepackage{epstopdf}
\usepackage{amsmath}

\def\apj{{ApJ}}
\def\apjs{{ApJS}}
\def\apjl{{ApJL}}

\def\aap{{A\&A}}
\def\pasp{{PASP}}

\def\mnras{{MNRAS}}
\def\nat{{Nature}}

\def\icarus{{Icarus}}

\shorttitle{Habitable Zone Statistics}
\shortauthors{Arthur D. Adams \& Stephen R. Kane}
\slugcomment{Accepted for publication in the Astronomical Journal}
\received{2016 January 20}
\accepted{2016 March 10}

\begin{document}

\title{Using {\it Kepler} Candidates to Examine the Properties of Habitable Zone Exoplanets}

\author{
  Arthur D. Adams\altaffilmark{1,2},
  Stephen R. Kane\altaffilmark{1}
}
\email{arthur.adams@yale.edu}
\altaffiltext{1}{Department of Physics \& Astronomy, San Francisco
  State University, 1600 Holloway Avenue, San Francisco, CA 94132,
  USA}
\altaffiltext{2}{Department of Astronomy, Yale University, 260 Whitney Avenue, New Haven,
  CT 06511, USA}

%%%%%%%%%%%%%%%%%%%%%%%%%%%%%%%%%%%%%%%%%%%%%%%%%%%%%%%%%%%%%%%%%%%%

\begin{abstract}

An analysis of the currently known exoplanets in the habitable zones (HZs) of their host stars is of interest in both the wake of the NASA Kepler mission and with prospects for expanding the known planet population through future ground- and space-based projects. In this paper we compare the empirical distributions of the properties of stellar systems with transiting planets to those with transiting HZ planets. This comparison includes two categories: confirmed/validated transiting planet systems, and Kepler planet and candidate planet systems. These two categories allow us to present quantitative analyses on both a conservative dataset of known planets and a more optimistic and numerous sample of Kepler candidates. Both are subject to similar instrumental and detection biases, and vetted against false positive detections. We examine whether the HZ distributions vary from the overall distributions in the Kepler sample with respect to planetary radius as well as stellar mass, effective temperature, and metallicity. We find that while the evidence is strongest in suggesting a difference between the size distributions of planets in the HZ and the overall size distribution, none of the statistical results provide strong empirical evidence for HZ planets or HZ planet-hosting stars to be significantly different from the full Kepler sample with respect to these properties.

\end{abstract}


%%%%%%%%%%%%%%%%%%%%%%%%%%%%%%%%%%%%%%%%%%%%%%%%%%%%%%%%%%%%%%%%%%%%

\section{Introduction}
\label{introduction}

The announcement of the validation of over 800 Kepler Objects of Interest (KOIs) through multiplicity validation \citep{lis14, roweetal2014} roughly doubles the existing number of confirmed/validated\footnote{We will refer to all known confirmed/validated exoplanets, including those detected by means other than the Kepler spacecraft, as ``C/V'' planets.} exoplanets in the catalog. At the current time more than 1900 confirmed or validated exoplanets are known \citep{wrightetal2011, akeson2013, mul15}. It is now estimated that Earth-sized planets or smaller occur more frequently than their Jupiter-sized or larger counterparts \citep{dressingcharbonneau2013}, with exoplanet detection missions like NASA's {\it Kepler space telescope} having expanded detections down to the regime of Earth-size planets orbiting stars at distances comparable to the Earth--Sun separation. In addition, planetary orbits are now found to exist in a wide range of eccentricities, with the range increasing for larger orbital distances \citep{kaneetal2012}.

We examine the planets detected by the Kepler mission, with particular attention paid to planets estimated to spend some fraction of their orbits within the Habitable Zones (HZs) of their host stars. With the unparalleled sensitivity of Kepler instrumentation the distribution of planets at this point is currently the most representative sample of the true exoplanet population within Kepler's detection capabilities. Several dozen confirmed planets and over 200 KOIs with well-determined Keplerian orbits (hereafter ``candidates'') are now estimated to spend the entirety of their orbits within the most optimistic formulation of the HZ. The boundaries of the HZ are defined in terms of the host stellar properties in Section \ref{hzb}, following the theoretical calculations of \citet{kopparapuetal2014}. Here we analyze the distributions of planets within and outside their systems' HZs, with respect to planetary size and orbital period, as well as host stellar mass and temperature.

Since there is an inherent bias in transit detection due to the geometric probability that a planet will transit along our line of sight, we use these transit probabilities to weight the distributions in an attempt to partly de-bias the sample. The results are presented in Section \ref{stats}. No assumptions are made other than assuming planetary orbits are randomly distributed in inclination angle, and that Kepler candidate planets represent true planet detections. Accordingly, our ``Kepler planets'' include both confirmed and candidate planets from Kepler. In doing so we rely on a completely empirical basis for comparing known transiting HZ planets with the superset of transiting planets. Using the Kepler planets we present statistical comparisons of the empirical distributions with respect to both properties of the planet or host star and whether the planet lies in the HZ of its host star. We expect the entire Kepler sample to represent currently the most complete sample of the nearby planet population, in terms of planet size, orbital period, and host stellar mass and temperature. By performing these statistical comparisons we hope to begin to answer whether the HZ planets reflect this distribution, or whether there are statistically significant differences with respect to properties of the planets or host stars.

Numerous works, including \citet{pet13} and \citet{for14}, have presented estimations of the occurrence fraction of Earth-sized planets in the HZ (often termed $\eta_\oplus$). This paper does not directly attempt a similar estimation, but rather examines what is currently available regarding the distribution of known planets in the HZ. The statistical techniques used here are limited to empirical inferences and do not extrapolate beyond the partial correction for geometric transit probability.

This study is motivated by results suggesting that planet radius exhibits dependence on semimajor axis. \citet{kanehinkel2013} use the major moons in our Solar System as analogs for compact planetary systems, and with Kepler-detected planets predict that the sizes of planets has at least a modest correlation with the size of orbit. \citet{mordasinietal2012} use the smaller maximum mass fraction of heavy elements in simulated protoplanetary disks interior to 1 AU compared with the mass fraction further out to conclude that a planet of a given mass will have a lower bulk density and thus larger radius the closer it lies to its host star.

The spectral type of planet-hosting stars has a demonstrated influence on the orbital evolution of planets \citep{bow10}, suggesting that such a dependence might lead to a difference in the planet rates in the HZ as a function of stellar mass and temperature. In addition, host stellar metallicity has been shown to be positively correlated with the occurrence rate of gas giant planets \citep{fis05, joh10, sou11}, while metallicity shows little to no correlation with the occurrence rate of planets with $R_{\textrm{P}} / R_\oplus < 4$ \citep{buc12}. It should be noted that \citet{fis05} were restricted by observational capacities at the time to so-called ``Hot Jupiters,'' with very short orbital distances, and even the more recent \citet{buc12} results relied on a sample that primarily contained planets with semimajor axes $<0.5$ AU. Therefore, we would like to examine firstly whether HZ planets, which extend the sample to larger orbital distances, might preferentially exist around stars whose metallicity distribution is different from that of planet hosts in general, by using the Kepler sample; and secondly whether in the ``small-radius'' regime (here $1 < R_{\textrm{P}} / R_\oplus < 5$ to limit ourselves to well-sampled regions of radius space) the radius distribution of HZ planets is different from those of higher metallicity stars (here super-solar), which should host giant planets more frequently than lower metallicity stars (sub-solar).

%%%%%%%%%%%%%%%%%%%%%%%%%%%%%%%%%%%%%%%%%%%%%%%%%%%%%%%%%%%%%%%%%%%%

\section{Methodology}
\label{calcs}

\subsection{HZ Boundaries}
\label{hzb}

The HZ is currently defined as the region surrounding a star where a terrestrial planet with an atmospheric pressure comparable to Earth can sustain liquid water on its surface. This analysis makes use of the optimistic HZ (OHZ) determinations as described in \citet{kopparapuetal2014}. This HZ is based on the 1-D climate models originally described in \citet{kastingetal1993}. These models assume that a planet would have an atmospheric pressure matching that of Earth, and water in some form present on the surface. The HZ is then determined by the distances the planet would need to lie from its host star such that the surface water would be in a liquid phase. The Recent Venus (RV) and Early Mars (EM) empirical limits, used to define the inner and outer bounds respectively of the OHZ, are first described in \citet{kopparapuetal2013}, and most recently in \citet{kopparapuetal2014}.

The percentage of orbital time a planet spends in the OHZ (denoted $t_{\textrm{OHZ}}/P$) can be calculated using a numerical method, described in \citet{kanegelino2012}. The star--planet separation given the best known orbit is sampled at 1000 equally time-spaced points, and at each is checked for whether it lies within the OHZ (hereafter simply ``HZ'').

%%%%%%%%%%%%%%%%%%%%%%%%%%%%%%%%%%%%%%%%%%%%%%%%%%%%%%%%%%%%%%%%%%%%

\subsection{Geometric Probability of Transit}
\label{transprob}

The probability for a planet to transit assuming isotropically oriented orbits can be approximated as \citep{kanevonbraun2008}

\begin{equation*}\label{eq:kanevonbraun2008}
P_t = \frac{\left( R_{\textrm{P}} + R_\star \right) \left( 1 + e \sin \omega \right)}{a \left( 1 - e^2 \right)}
\end{equation*}
where $a$ is the semimajor axis, $e$ the orbital eccentricity, and $\omega$ the argument of periastron. All distributions except for orbital eccentricity are weighted by the inverse of the transit probability. For all further discussions except those concerning orbital eccentricity, ``C/V'' will only refer to those confirmed/validated planets which have transit detections.

\subsection{Removing Outliers from the Candidates}
The current database of objects listed as candidate planets \citep{mul15, row15} contains some caveats that have the potential to skew the distribution in unphysical ways. We impose two cuts to the data which are motivated both by the detection capabilities of the Kepler spacecraft and peculiarities in the Kepler data processing pipeline:
\begin{enumerate}
\item $R_{\textrm{P}} / R_\star < 0.19$. \citet{mul15} note (see Section 7) that certain factors in data processing may lead to planet radii being drastically overestimated, including, but not limited to, misclassification of main-sequence stars as evolved stars. The specific value for this cut comes from the maximum planet--star radius ratio found among all confirmed exoplanets in the NASA Exoplanet Archive.

\item Orbital period $<1$500 days. There are a handful of listed planet candidates with orbital periods longer than the lifetime of the original Kepler mission. We impose a simple cut to exclude those candidates which could not have had the possibility of at least two observed transits.
\end{enumerate}

\subsection{Statistical Methods}
For each sample we generate 10,000 simulations of planet size, stellar mass, and stellar temperature distributions, each with a size equal to that of the HZ sample. The weighted data are used as discrete probability distributions for the simulations. The full and HZ-only simulations are paired and compared in pairs statistically using an Anderson--Darling test, with the null hypothesis each pair of simulations is drawn from the same underlying property distribution. From this we obtain a distribution of $p$-values, which has some proportion of simulations with the $p$-value below some chosen value: for this paper, $p<0.01$ and $p<0.05$ are considered. This $p$-value represents the probability that a HZ subset could be drawn by chance from the full sample and be at least as different from the full sample as what we drew. The difference here is between the cumulative distribution functions of the planetary or stellar parameter under consideration. We use this proportion to estimate how strongly the evidence is against the null hypothesis, with a higher proportion providing a stronger case of a difference between the overall and HZ distributions with respect to that parameter.

%%%%%%%%%%%%%%%%%%%%%%%%%%%%%%%%%%%%%%%%%%%%%%%%%%%%%%%%%%%%%%%%%%%%

\section{Discussion}
\label{stats}

\begin{figure*}
  \begin{center}
    \includegraphics[angle=0,width=16.4cm]{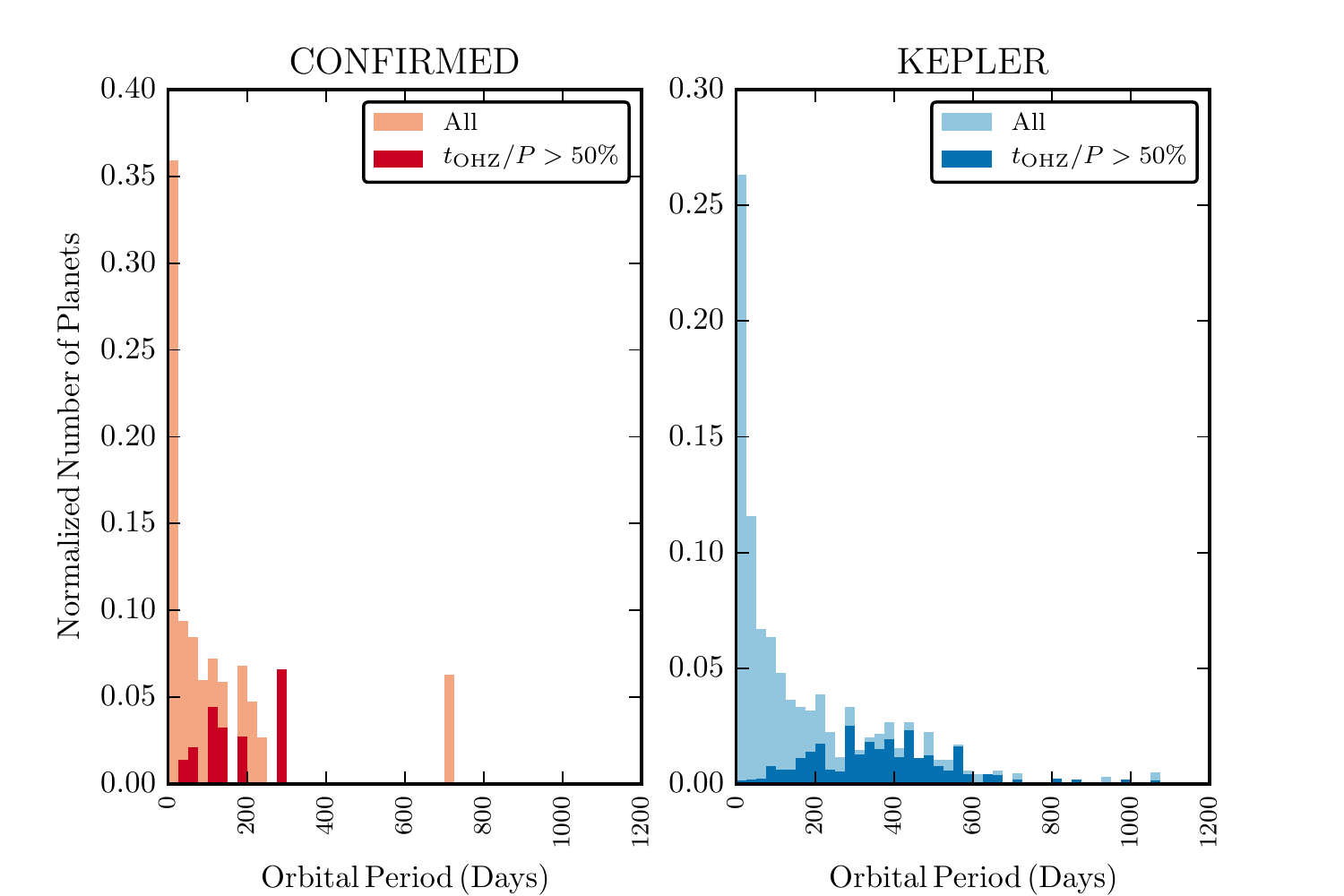}
  \end{center}
  \caption{The distribution of transiting C/V and Kepler planets with respect to orbital period. The data are shaded according to the percentage of time each planet spends in its HZ.}
  \label{Fig:OrbitHists}
\end{figure*}

\begin{figure*}
  \begin{center}
    \includegraphics[angle=0,width=16.4cm]{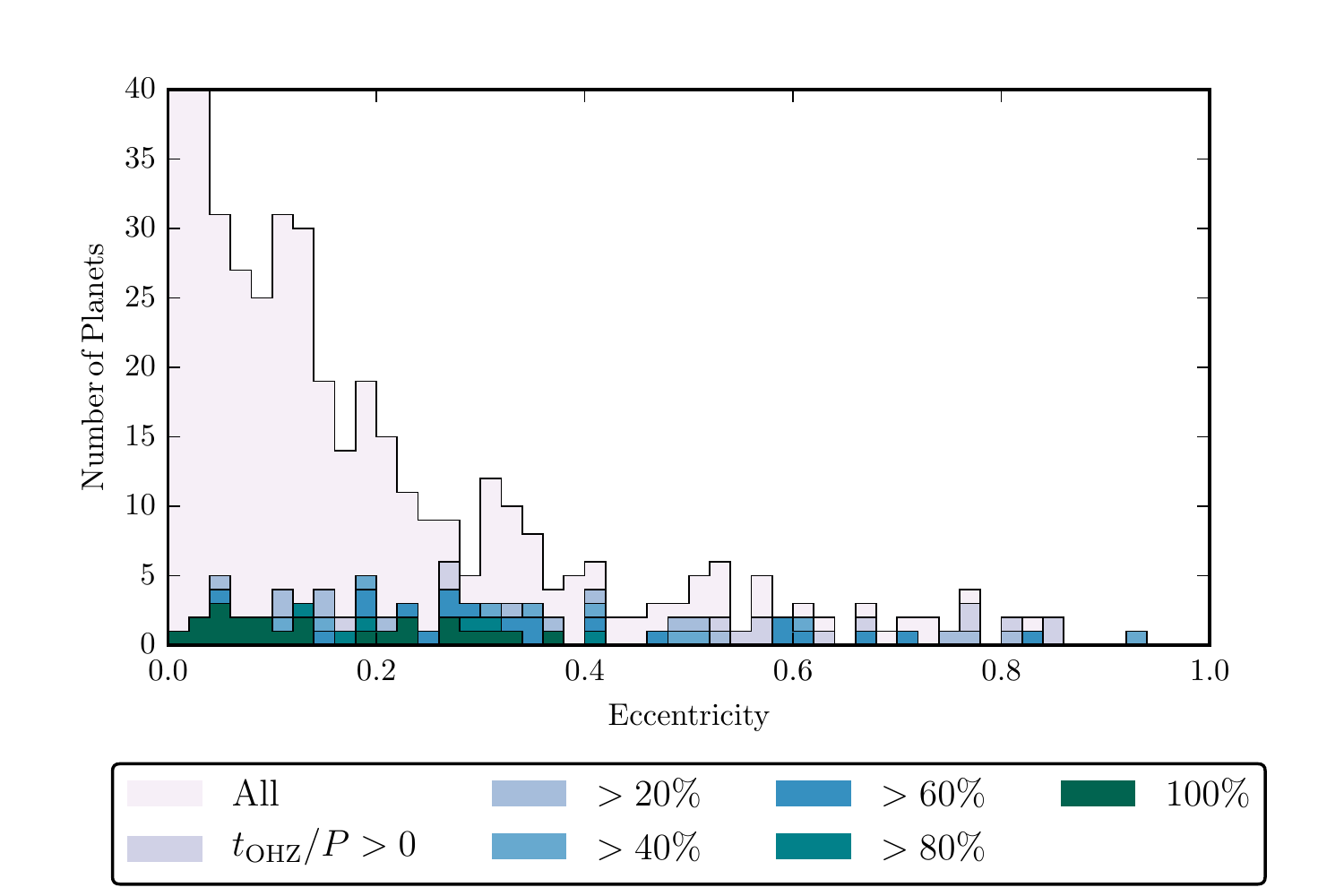}
  \end{center}
  \caption{The distribution of all known C/V exoplanets with respect to orbital eccentricity. The data are color-coded according to the percentage of time each planet spends in its HZ.}
  \label{Fig:EccHist}
\end{figure*}

Figures \ref{Fig:OrbitHists}, \ref{Fig:StellarHists}, \ref{Fig:StellarTeffHists}, and \ref{Fig:PlanetHists} show the transit probability-weighted distributions of both C/V transiting planets and Kepler planets, binned according to selected system properties and shaded according to $t_{\textrm{OHZ}}/P$. Table \ref{table:planets} gives the number of planets in each distribution, along with the weighted means and medians for each parameter. ``HZ planets'' here are those planets which spend at least half their determined orbital time within the HZ. Most of the Kepler planets do not have measured eccentricities; in this case, HZ planets will be ones whose measured semimajor axis lies within the HZ.

Kepler candidate planets may only be confirmed or validated through methods such as RV follow-up detection, transit timing methods \citep{fabryckyetal2012}, the BLENDER analysis to rule out false positives from eclipsing binaries or external light sources \citep{tor11}, or multiplicity validations \citep{lissaueretal2011, lissaueretal2012, roweetal2014}. As such, only candidates which have a transit and an applicable secondary detection can be confirmed or validated. The resulting sample of C/V planets from Kepler are expected to contain observational biases. For this reason we choose to restrict our statistical comparisons to the Kepler planets, whose results are shown in the right panels of the figures mentioned above.

\subsection{Properties of the Orbits}
\label{stats:orbits}
We examine the occurrence rate of HZ planets with respect to orbital period (Figure \ref{Fig:OrbitHists}). The planets in the C/V sample are heavily skewed toward short orbital periods, as to be expected by the observational bias inherent both in RV and transit detections. We see very few HZ planets, and all orbit within one Earth year. RV-detected planets are preferentially closer in to their host stars, while Kepler planets are limited primarily by the time extent of observations. The latter is not a sensitivity bias but rather a limitation from the length of the mission. Therefore we expect the distribution of the Kepler planets to be more representative of the exoplanet population (after transit probability weighting) out to the periods possible with the mission. From the Kepler sample we see that HZ planets span nearly the entire range of observed orbital periods, with a mean very close to one Earth year. This is also consistent with the target selection of the Kepler mission, focusing on Solar-type stars whose HZs in period space would be close to that of Earth.

The exploration of theoretical habitable conditions on planets in orbits which are both highly eccentric and largely within the HZ can be found in the literature \citep{wil02, kanegelino2012-2}, and the existence of such planets in the presently known population warrants further consideration of this phenomena quite unlike what is observed in the Solar System. The distribution of orbital eccentricities for all C/V planets (Figure \ref{Fig:EccHist}) shows a wide range of eccentricities for planets which are estimated to have considerable $t_{\textrm{OHZ}}/P$ values. In particular, 19 confirmed exoplanets with $e \geq 0.3$ have $t_{\textrm{OHZ}}/\P \geq 50\%$. Many of the Kepler planets do not have sufficient information about their orbits to put constraints on their eccentricities; where they are undetermined, circular orbits are assumed.

%%%%%%%%%%%%%%%%%%%%%%%%%%%%%%%%%%%%%%%%%%%%%%%%%%%%%%%%%

\subsection{Properties of the Host Stars}
\label{stats:stars}

\subsubsection{Stellar Mass and Effective Temperature}
\begin{figure*}
  \begin{center}
      \includegraphics[angle=0,width=16.4cm]{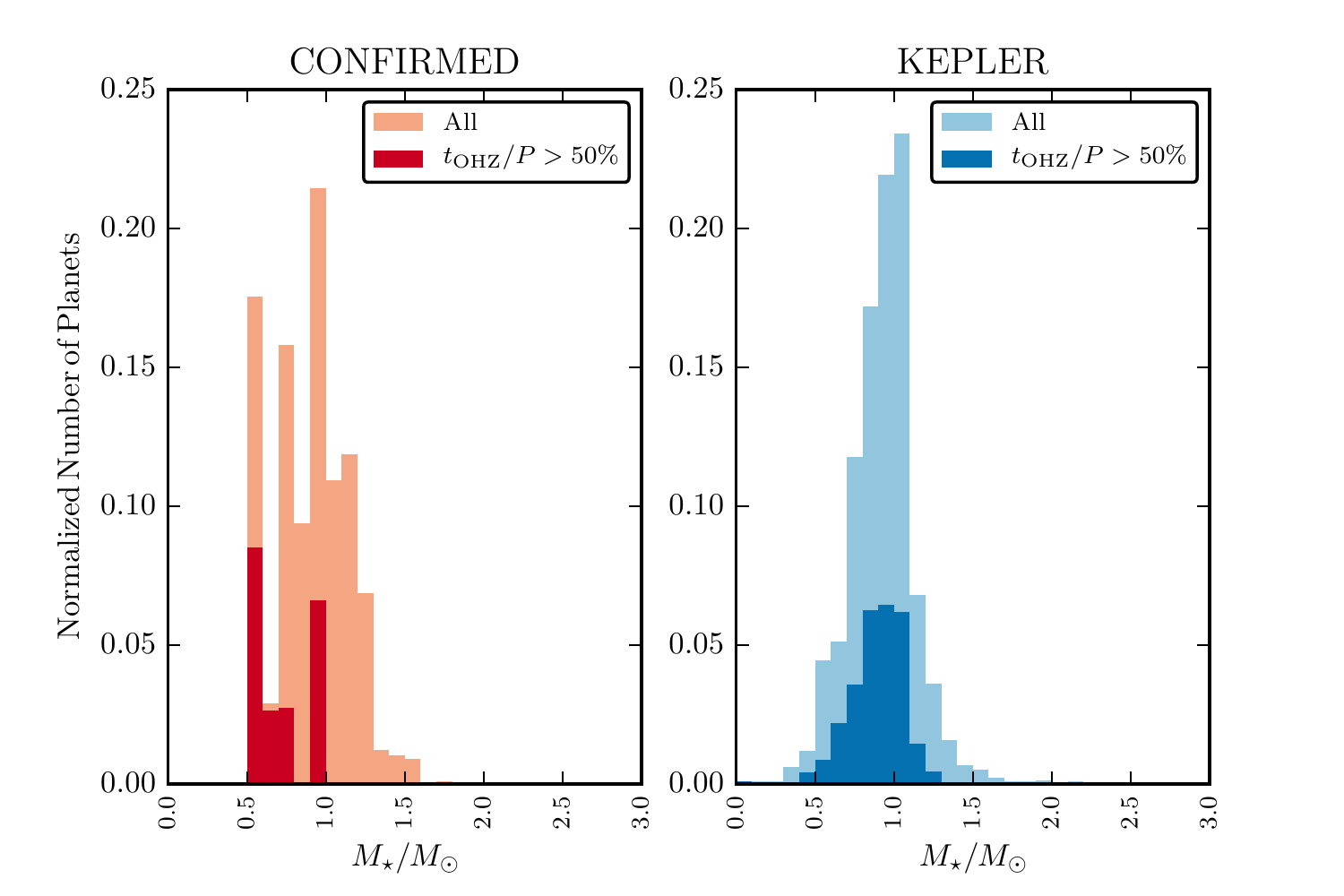}
  \end{center}
  \caption{The distribution of transiting C/V and Kepler planets with respect to host stellar mass. The data are shaded according to the percentage of time each planet spends in its HZ.}
  \label{Fig:StellarHists}
\end{figure*}

\begin{figure*}
  \begin{center}
      \includegraphics[angle=0,width=16.4cm]{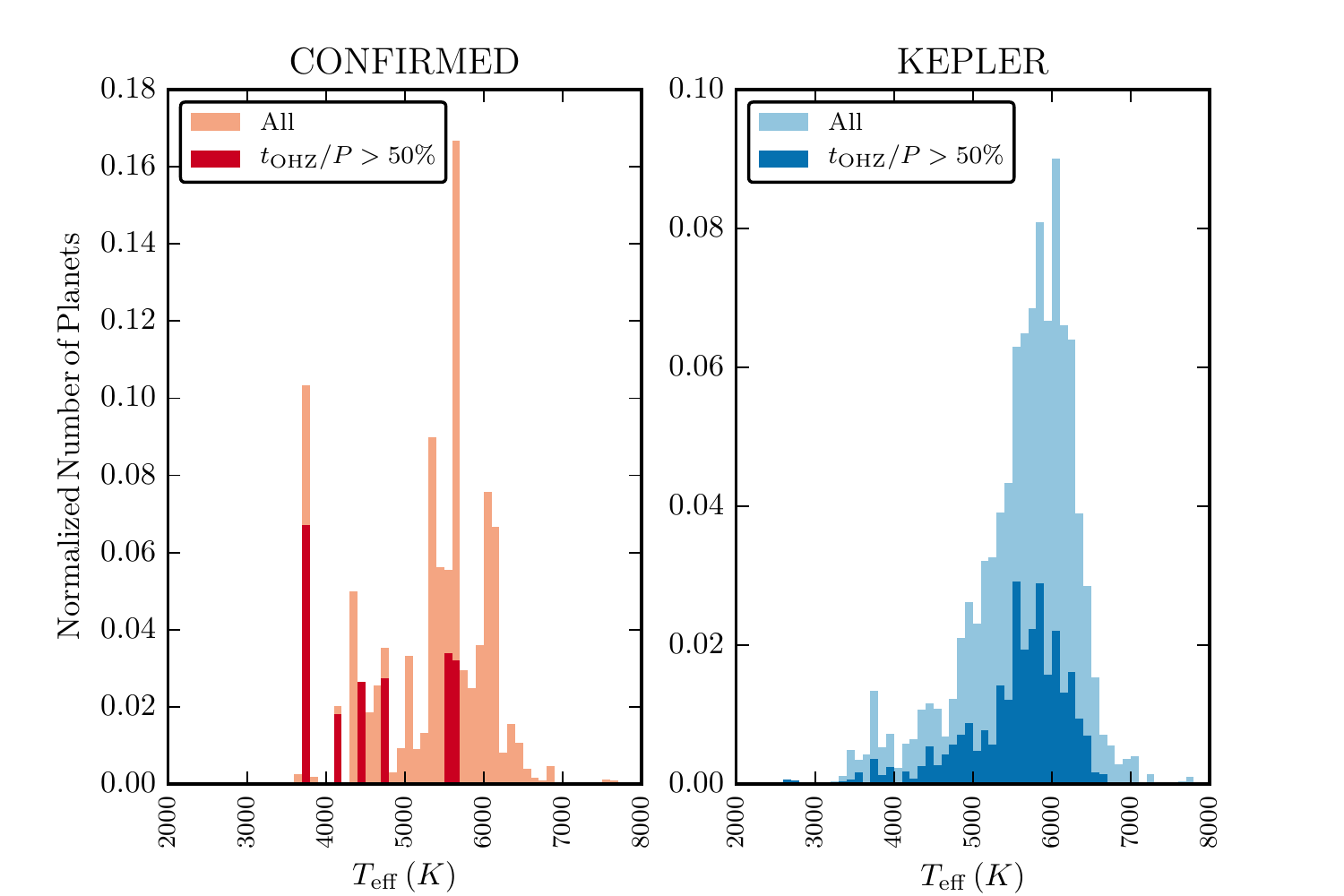}
  \end{center}
  \caption{The distribution of transiting C/V and Kepler planets with respect to host stellar effective temperature. The data are shaded according to the percentage of time each planet spends in its HZ.}
  \label{Fig:StellarTeffHists}
\end{figure*}

We consider the distribution of HZ planets with respect to host stellar mass and effective temperature (Figures \ref{Fig:StellarHists} and \ref{Fig:StellarTeffHists}). The C/V distribution has mean values that are sub-solar, while the Kepler distribution means, despite also being sub-solar, are roughly consistent within uncertainties with solar values. We attribute the latter to the selection criteria of the Kepler mission, which contributes to both but dominates most the Kepler sample. Since Kepler relies exclusively on photometry to detect planets, its flux limitations extend to much fainter apparent magnitudes than RV survey stars. Therefore we expect the ability to detect transits around faint stars (which in a volume-limited sample corresponds to later-type stars), despite the focus of Kepler on solar analogs.

\begin{figure*}
  \begin{center}
    \includegraphics[angle=0,width=16.4cm]{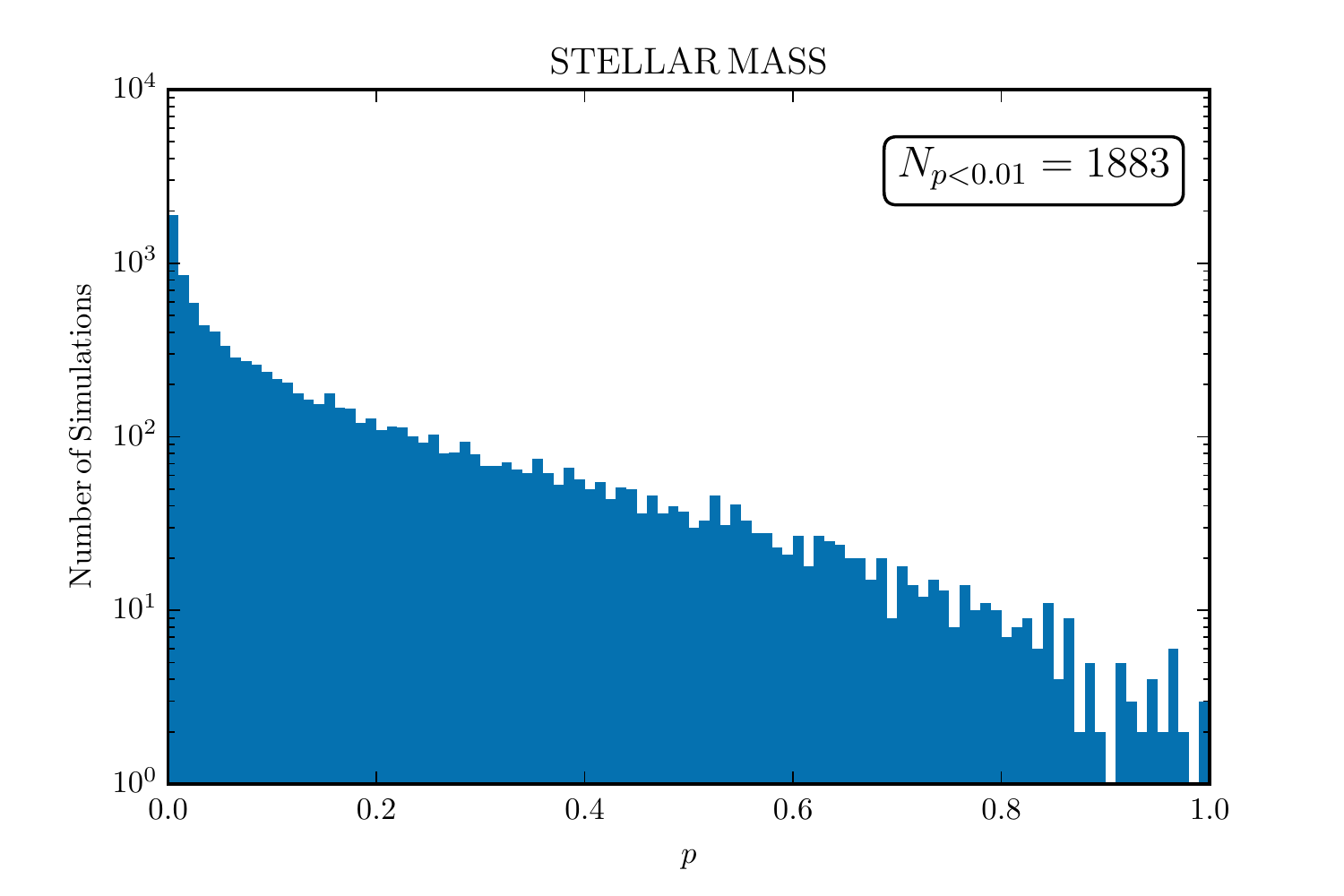}
  \end{center}
  \caption{The distribution of $p$-values from two-sample A--D tests between random samples drawn from the entire distribution of stellar masses for Kepler systems, and a sample drawn from the distribution of masses for only those Kepler systems with planets in the HZ. Each sample is of the same size as the total number of Kepler HZ planets.}
  \label{Fig:smass_pvals}
\end{figure*}

\begin{figure*}
  \begin{center}
    \includegraphics[angle=0,width=16.4cm]{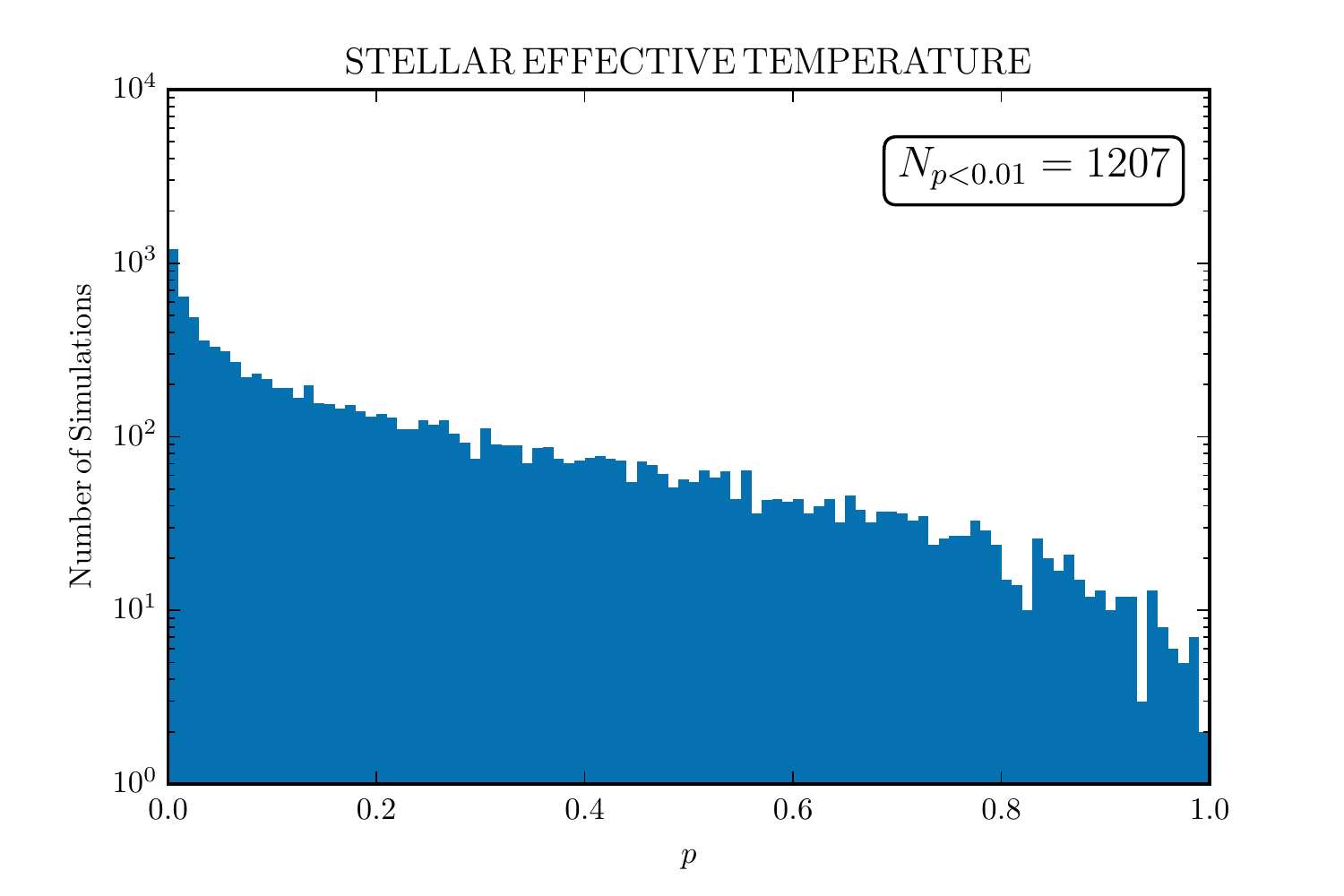}
  \end{center}
  \caption{The distribution of $p$-values from two-sample A--D tests between random samples drawn from the entire distribution of stellar effective temperatures for Kepler systems, and a sample drawn from the distribution of temperatures for only those Kepler systems with planets in the HZ. Each sample is of the same size as the total number of Kepler HZ planets.}
  \label{Fig:steff_pvals}
\end{figure*}

There is little evidence for differences between the distributions of masses of Kepler planet hosts and HZ planet hosts (Figure \ref{Fig:smass_pvals}). Only 1883 out of our 10,000 A--D tests return $p<0.01$. Even weaker is the evidence of a difference in the effective temperature distributions, with only 1207 of the tests returning $p<0.01$ (Figure \ref{Fig:steff_pvals}). Based on these results there is insufficient evidence to suggest that stars hosting HZ planets differ in mass and temperature significantly from planet-hosting stars in general. Target selection for Kepler stars and the biases therein might overwhelm any astrophysically motivated differences. This ambiguity is likely exacerbated by the small sample size of currently known HZ planets.

\subsubsection{Stellar Metallicity}\label{sec:metal}
\begin{figure*}
  \begin{center}
      \includegraphics[angle=0,width=16.4cm]{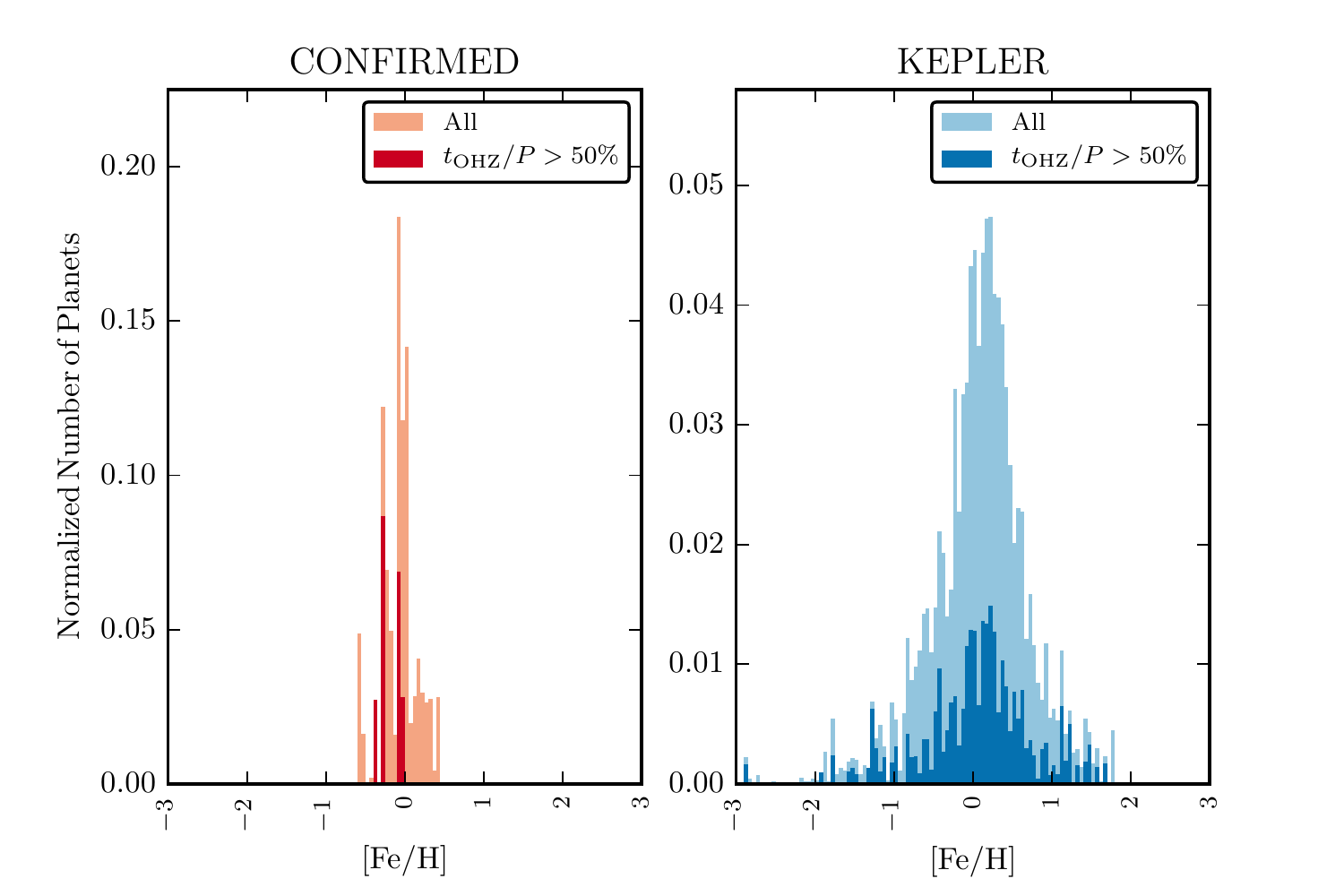}
  \end{center}
  \caption{The distribution of transiting C/V and Kepler planets with respect to host stellar metallicity. The data are shaded according to the percentage of time each planet spends in its HZ.}
  \label{Fig:StellarFeHHists}
\end{figure*}

\begin{figure*}
  \begin{center}
    \includegraphics[angle=0,width=16.4cm]{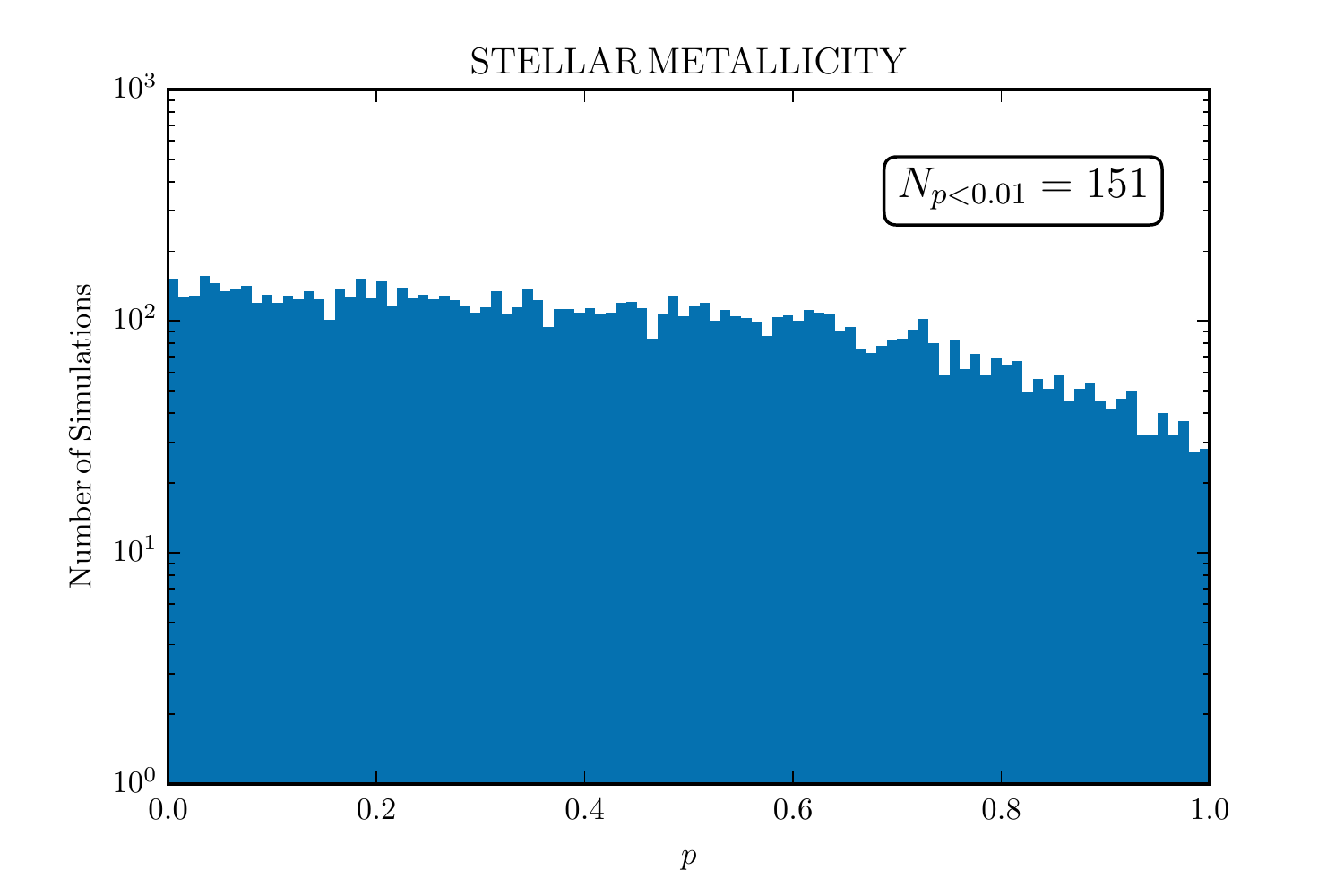}
  \end{center}
  \caption{The distribution of $p$-values from two-sample A--D tests between random samples drawn from the entire distribution of stellar metallicities for Kepler systems, and a sample drawn from the distribution of metallicities for only those Kepler systems with planets in the HZ. Each sample is of the same size as the total number of Kepler HZ planets.}
  \label{Fig:stmet_pvals}
\end{figure*}

While many of the Kepler stars have metallicities listed in the {\it Kepler Input Catalog} (KIC), the reliability of individual metallicities is questionable. \citet{don14} compare KIC metallicities for a sample of Kepler stars with those obtained with spectroscopy and find a systematic offset
\begin{equation*}
\left[\mathrm{Fe}/\mathrm{H}\right]_{\mathrm{KIC}} = \left(-0.203 \pm 0.002\right) + \left(0.434 \pm 0.011\right) \left[\mathrm{Fe}/\mathrm{H}\right]_{\mathrm{LAMOST}}
\end{equation*}
with a resulting scatter of roughly 0.25 dex in the binned data. \citet{don14} caution that the resulting ``true'' metallicities using an inversion of this relation will be unreliable due to an estimated $\sim 0.6$ dex scatter. Nevertheless, we incorporate these values and associated uncertainties into our analysis (Figure \ref{Fig:StellarFeHHists}), allowing each random sample to vary uniformly within the uncertainty range for individual metallicities. The uncertainties overwhelm any potential differences in the distributions, with the evidence for a difference nonexistent: only 151 of 10,000 tests return $p<0.01$ (Figure \ref{Fig:stmet_pvals}). This lack of evidence persists even when the uncertainties are not considered. The lack of evidence for a metallicity correlation for HZ hosts is consistent with the findings of \citet{fis05} and \citet{buc12}, and is further corroborated when one compares the radius distribution of small-radius planets ($1 < R_{\textrm{P}}/R_\oplus < 5$, see Section \ref{stats:planets}) around metal-rich hosts, compared with those in the HZ.

%%%%%%%%%%%%%%%%%%%%%%%%%%%%%%%%%%%%%%%%%%%%%%%%%%%%%%%%%

\subsection{Properties of the Planets}
\label{stats:planets}
\begin{figure*}
  \begin{center}
    \includegraphics[width=16.4cm]{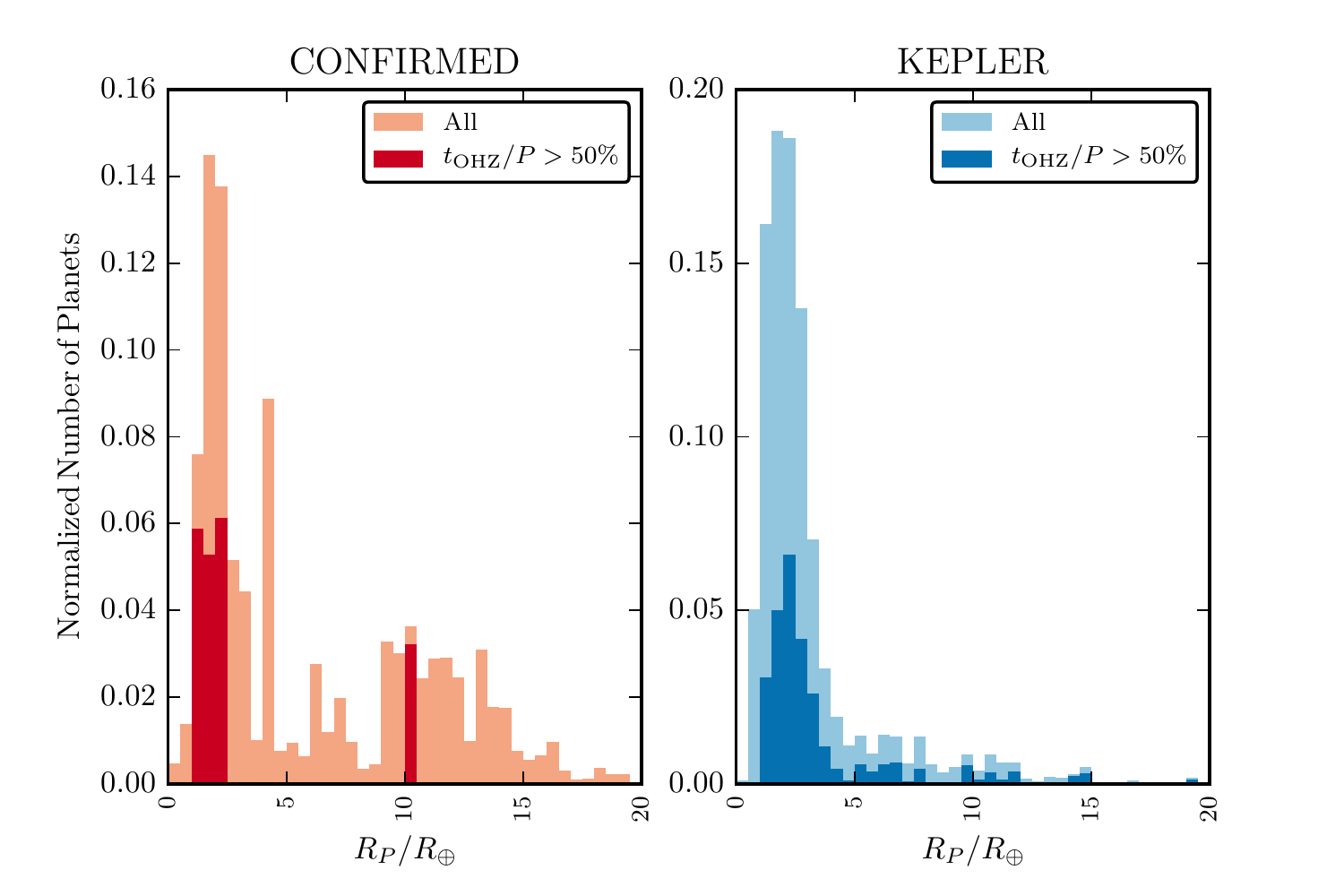}
  \end{center}
  \caption{The distribution of transiting C/V and candidate planets with respect to planetary radius. The data are color-coded according to the percentage of time each planet spends in its HZ.}
  \label{Fig:PlanetHists}
\end{figure*}

\begin{figure*}
  \begin{center}
   \begin{tabular}{cc}
     \includegraphics[width=8.2cm]{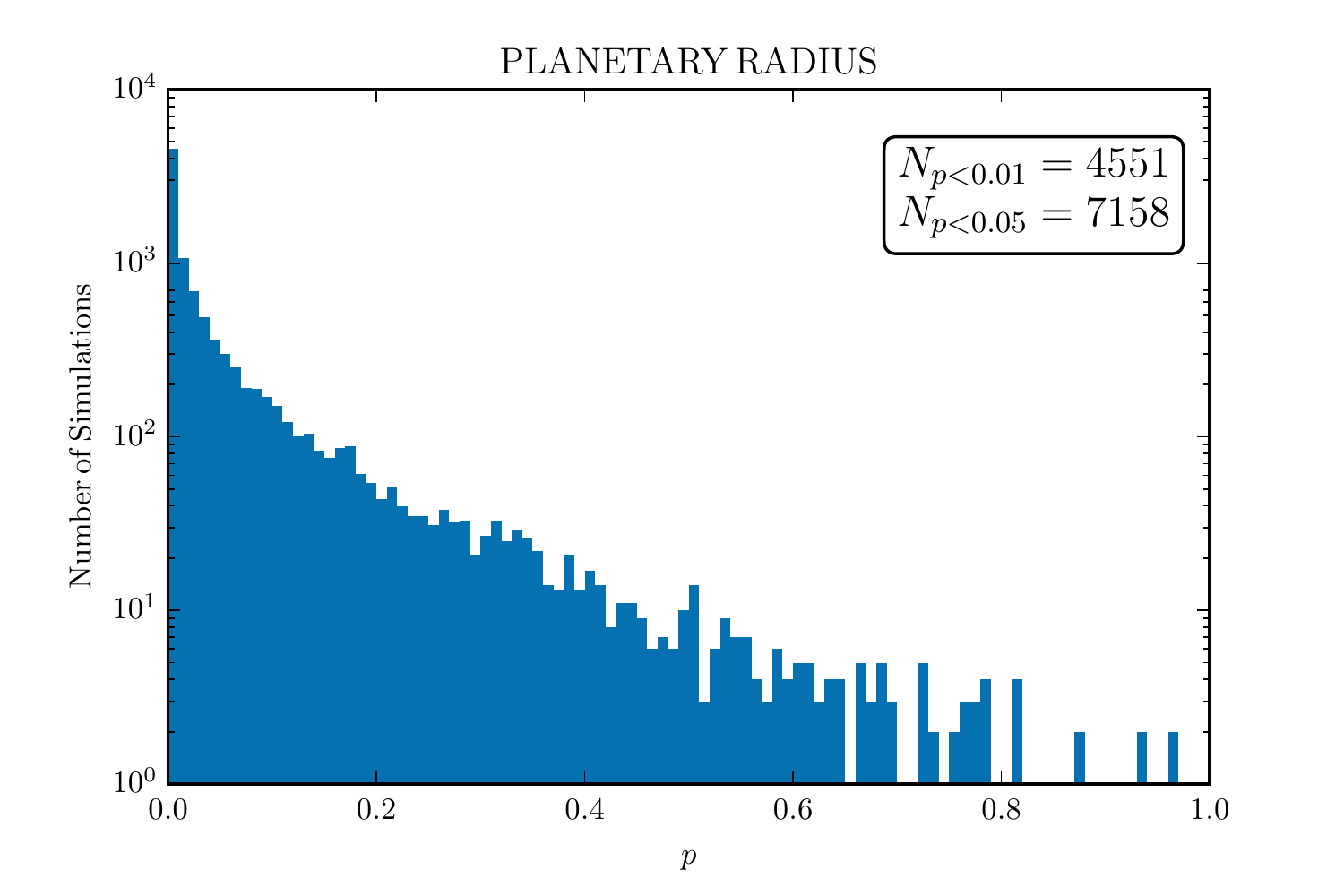} &
     \includegraphics[width=8.2cm]{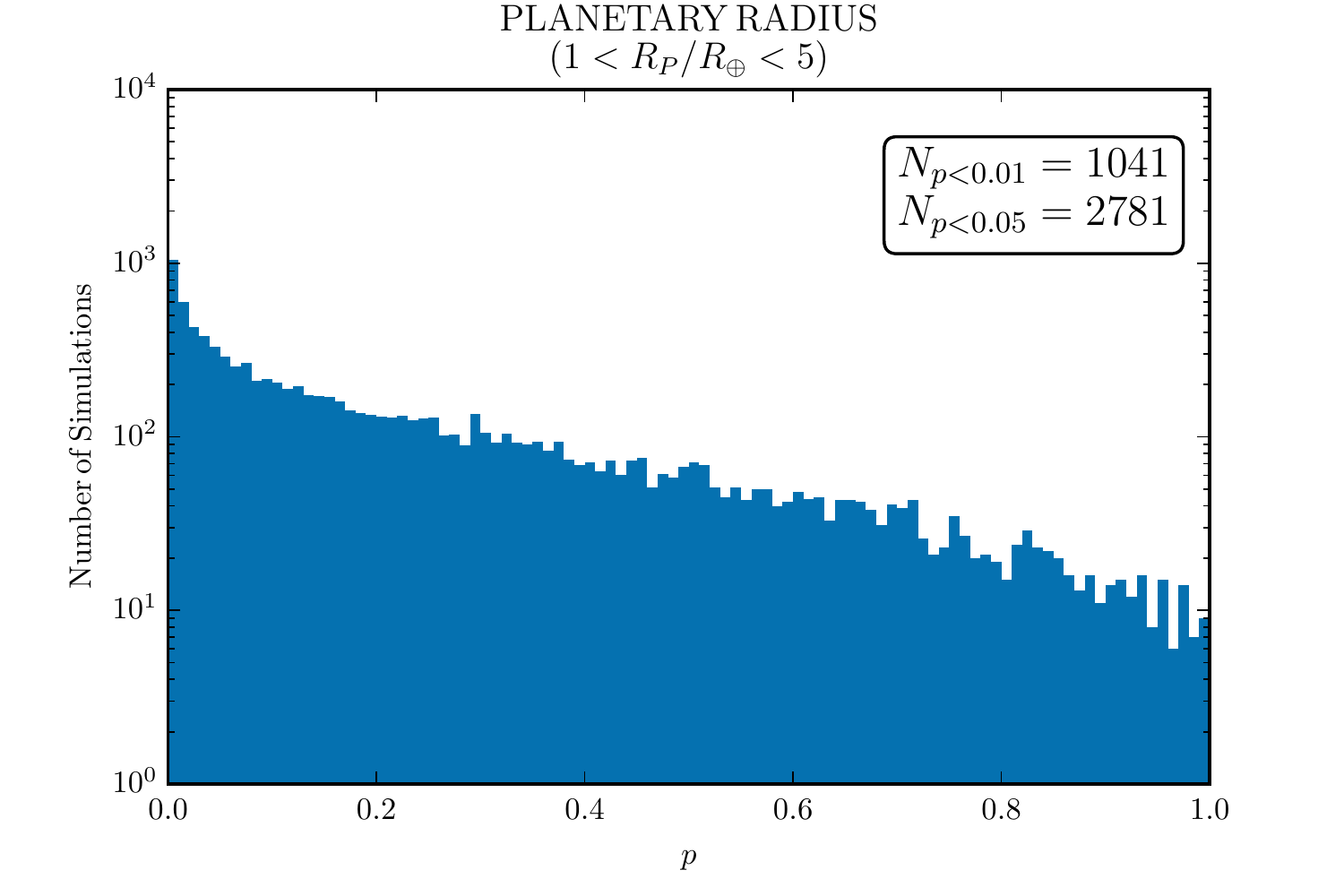}
    \end{tabular}
  \end{center}
  \caption{The distributions of $p$-values from two-sample A--D tests between random samples drawn from the entire distribution of planetary radii for Kepler systems, and a sample drawn from the distribution of radii for only those Kepler planets in the HZ. Each sample is of the same size as the total number of Kepler HZ planets. Left: the distribution for all planets meeting our imposed criterion $R_{\textrm{P}} / R_\star < 0.19$. Right: the distribution restricting the sample to planets with $1 < R_{\textrm{P}} / R_\oplus < 5$. At the $p<0.05$ level well more than half of the tests return a statistically significant difference in the radius distribution of Kepler planets compared with Kepler planets in the HZ. The number of statistically significant results decreases drastically when the radius cut is imposed, demonstrating that the distributions of planets within the peaks of either sample are not significantly different.}
  \label{Fig:plrad_pvals}
\end{figure*}

We lastly examine the distributions of planetary radii (Figure \ref{Fig:PlanetHists}). We see that in both the overall and HZ distributions for the Kepler planets the mean size is between 2 and 3$R_\oplus$, with tails extending to Jupiter-sized planets and above. The C/V sample contains many Kepler planets, with existing RV planets outside the HZ greatly biased toward more massive/larger planets.  From the results of the statistical tests (Figure \ref{Fig:plrad_pvals}) there initially seems to be evidence one can reject the null hypothesis for a comparison of the candidate overall and HZ samples at $p<0.05$, with well over half (7158 of 10,000) of the tests meeting the criterion. However, this result is likely due to the persistence of observational biases in the low-radius and mass regime, since weighting cannot debias regions where no instances of planets exist (in this case, in the HZ subset). As such, it is not clear that planets have a significantly different size distribution solely on the basis of whether they lie in the HZs of their host stars. To examine whether the tails skew the results of our statistical tests, we repeated the comparison restricting ourselves to planets in the peak of the distributions: $1 < R_P/R_\oplus < 5$. Then the number of significant results drops to only 2,781 for $p<0.05$, suggesting that the few outliers and lack of sub-Earth HZ planets have a major effect on our statistical analysis.

\begin{figure*}
  \begin{center}
    \includegraphics[width=16.4cm]{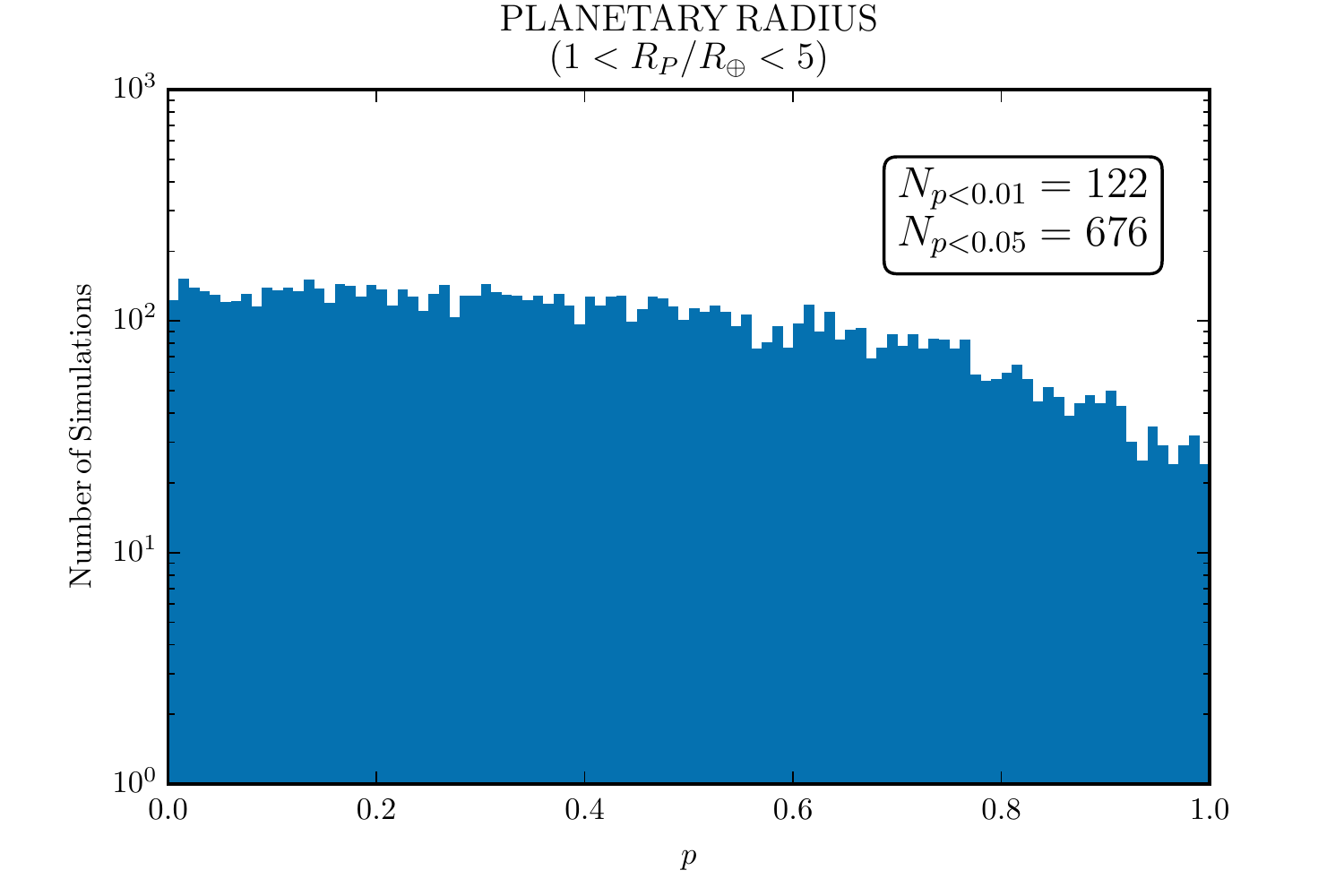}
  \end{center}
  \caption{The distributions of $p$-values from two-sample A--D tests between random samples drawn from planetary radii for Kepler systems with super-solar stellar metallicities, and a sample drawn from the distribution of HZ planet radii. In this case, we restrict ourselves to planets with $1 < R_{\textrm{P}} / R_\oplus < 5$. We conclude there is no evidence for significant differences between the distribution of small planets for high-metallicity planet hosts and the distribution for small HZ planets.}
  \label{Fig:supe_planets}
\end{figure*}

To test whether metal-rich planet hosts might have statistically higher planet radii for the Earth-sized through Neptune-sized regime, we restrict ourselves to planets between 1 and 5 Earth radii $\left(1 < R_{\textrm{P}}/R_\oplus < 5\right)$ and perform a statistical comparison of planet radii around metal-rich hosts with those of HZ planets (Figure \ref{Fig:supe_planets}). As noted in Section \ref{sec:metal}, the lack of metallicity correlation is present here as well. The Kepler sample lends no evidence that high-metallicity systems, which tend to host giant planets more frequently but otherwise have a similar distribution of planet radii \citep{fis05,buc12}, have a different occurrence rate of super-Earths/sub-Neptunes from that of the Kepler HZ planets.

%%%%%%%%%%%%%%%%%%%%%%%%%%%%%%%%%%%%%%%%%%%%%%%%%%%%%%%%%

\begin{deluxetable}{c r | c c c c c c c c c c}
  \rotate
  \tabletypesize{\scriptsize}
  \tablecolumns{7}
  \tablewidth{0pc}
  \tablecaption{\label{table:sma_stats}Mean and Median Values for System Parameters}
  \tablehead{
    \colhead{$t_{\textrm{OHZ}}/P$} 			&
    \colhead{$N_{\textrm{pl}}$\tablenotemark{a}} 									&
    \multicolumn{2}{c}{Orbital Period (Days)}			&
    \multicolumn{2}{c}{ $M_\star / M_\odot$}	&
    \multicolumn{2}{c}{$T_{\mathrm{eff},\star}$ (K)}		&
    \multicolumn{2}{c}{$\left[\mathrm{Fe}/\mathrm{H}\right]$}		&
    \multicolumn{2}{c}{Planetary Radius $\left(R_\oplus\right)$}\\
    \colhead{}									&
    \colhead{}									&										
    \colhead{Mean} &
    \colhead{Median} &
    \colhead{Mean} &
    \colhead{Median} &
    \colhead{Mean} &
    \colhead{Median} &
    \colhead{Mean} &
    \colhead{Median} &
    \colhead{Mean} &
    \colhead{Median} \\
  }
  \startdata
  \sidehead{Kepler}
 0\% -- 100\% & 3343 & $178.263\pm0.005$ & $95.907\pm0.001$ & $0.93^{+0.14}_{-0.09}$ & $0.95^{+0.13}_{-0.08}$ & $5632^{+157}_{-137}$ & $5772\tablenotemark{b}^{+180}_{-146}$ & $0.06\pm0.58$ & $0.12\pm0.60$ & $3.24^{+1.22}_{-0.43}$ & $2.26^{+0.96}_{-0.35}$ \\
 $>50\%$ & 197 & $365.25\pm0.01$ & $358.818\pm0.003$ & $0.89^{+0.12}_{-0.07}$ & $0.91^{+0.12}_{-0.08}$ & $5555^{+160}_{-124}$ & $5696^{+166}_{-162}$ & $0.05\pm0.58$ & $0.12\pm0.60$ & $3.62^{+1.14}_{-0.40}$ & $2.42\pm0.23$ \\
   \sidehead{C/V}
 0\% -- 100\% & 203 & $124.254\pm0.001$ & $64.0020\pm0.0007$ & $0.90\pm0.05$ & $0.92\pm0.03$ & $5268^{+92}_{-91}$ & $5500\pm100$ & $-0.08\pm0.10$ & $-0.08\pm0.03$ & $6.15^{+0.47}_{-0.43}$ & $4.16^{+0.19}_{-0.16}$ \\
 $>50\%$ & 8 & $168.565\pm0.002$ & $129.944\pm0.001$ & $0.71\pm0.04$ & $0.61\pm0.03$ & $4584\pm90$ & $4402\pm100$ & $-0.19\pm0.12$ & $-0.26\pm0.12$ & $3.10^{+0.23}_{-0.24}$ & $1.86^{+0.24}_{-0.19}$ \\
 \enddata
 \label{table:planets}
 \tablenotetext{a}{From the NASA Exoplanet Archive, as of 2015 September 10.}
 \tablenotetext{b}{Note: For planets without well-determined stellar properties, effective temperature is set to the solar value \citep{mul15}.}
\end{deluxetable}

%%%%%%%%%%%%%%%%%%%%%%%%%%%%%%%%%%%%%%%%%%%%%%%%%%%%%%%%%%%%%%%%%%%%

\section{Conclusions}
\label{concs}

With the number of confirmed exoplanets surpassing 1500 and a candidate planet sample from Kepler in excess of 3300, we now know of a considerable number of planets which spend a large fraction of their orbits within the HZ. We have presented the distribution of said planets and planet candidates as a function of properties of both the planets and their host sizes, and explored the extent to which the most current sample of planets and planet candidates has expanded the range of observed planet sizes, orbits, and planet-hosting stars. Despite the wealth of knowledge from Kepler, there are still regimes of planet detection that are yet unreachable, and only when these can be probed will we have a better understanding of the true HZ planet population.

Several salient caveats exist with the current statistical analysis. Firstly is that the true false-positive rate for the Kepler planet candidates is still not known precisely, and is still evolving with the improvement of verification methods. It has been suggested \citep{san12, fre13} that the rate of false positives may depend on orbital period, which in part might be due to the rotation rate of the Kepler spacecraft. Second, stellar uncertainties are still higher for radii than for effective temperature \citep{akeson2013}, with many radial uncertainties being a large fraction of the determined radius. Since the determined size of the planet is directly dependent on the determined stellar size, uncertainties will propagate to the measured planet size, which from studies such as \citet{lopezfortney2014} is a proxy for determining the bulk composition of planets within a factor of 2 of the size of Earth, as well as the boundaries of the HZ \citep{kane2014}.

Study of the types of planets and stars that observations to date have discovered is beneficial for the refinement of host stellar target selection. While the mean stellar size and temperature for HZ hosts are centered roughly at solar values, HZ host stars in the known sample have radii down to $\approx0.5 R_\odot$. With the announcement of Kepler-186 f \citep{quintanaetal2014}, an Earth-like planet orbiting an M-dwarf, along with the known higher probability of transit for HZ planets around later-type stars, there is motivation to search for HZ planets around K- and M-dwarfs. Additionally, the presence of 19 confirmed exoplanets with large orbital eccentricities ($\geq 0.3$) and at least $\geq 50 \%$ of their orbital time in the HZ warrants further study of planets in eccentric orbits with respect to surface conditions on the planet \citep{wil02, kanegelino2012-2}, especially since the eccentricity distribution of the Kepler candidates is as of yet undetermined but estimated to follow the distribution seen in RV-detected planets \citep{kaneetal2012}.

Numerous works have been written concerning the robustness of planet formation theories with the wealth of Kepler data \citep{rei12, han13, swi13, cha14}. \citet{chi13} posit that with an in-situ formation theory, early-type main-sequence stars should generally lack rocky super-Earths at distances analogous to the inner region of our Solar System $\left( \sim 0.1 - 10 \, \mathrm{AU} \right)$. Consistent with this prediction, the hottest star in the candidate sample to host a HZ planet with $R<1.5R_\oplus$, Kepler-371, has an effective temperature of 6038 K, placing it at the cool end of the F dwarfs. Future work will explore the consistency of the Kepler data more quantitatively with respect to theories of planet formation, with attention paid to the observable effects on the properties of planets in the HZ of their host stars.

%%%%%%%%%%%%%%%%%%%%%%%%%%%%%%%%%%%%%%%%%%%%%%%%%%%%%%%%%%%%%%%%%%%%

\section*{Acknowledgements}

This research has made use of the following online resources: the
Exoplanet Orbit Database and the Exoplanet Data Explorer at
exoplanets.org, the Habitable Zone Gallery at hzgallery.org, and the NASA Exoplanet Archive,
which is operated by the California Institute of Technology,
under contract with the National Aeronautics and Space Administration under the Exoplanet Exploration Program.

This research has benefited from generous funding from the Northern California ARCS Foundation, as well as the Gruber Foundation at Yale University.

Additionally, we would like to acknowledge Professors Andrew Barron and Huibin Zhou of the Yale Statistics Department, whose insightful comments contributed to a refinement of the analysis in this paper.

%%%%%%%%%%%%%%%%%%%%%%%%%%%%%%%%%%%%%%%%%%%%%%%%%%%%%%%%%%%%%%%%%%%%


\begin{thebibliography}{}

\bibitem[Akeson et al.(2013)]{akeson2013} Akeson, R.~L., Chen, X., 
Ciardi, D., et al.\ 2013, \pasp, 125, 989 

\bibitem[Bowler et al.(2010)]{bow10} Bowler, B.~P., Johnson, 
J.~A., Marcy, G.~W., et al.\ 2010, \apj, 709, 396 

\bibitem[Buchhave et al.(2012)]{buc12} Buchhave, L.~A., 
Latham, D.~W., Johansen, A., et al.\ 2012, \nat, 486, 375 

\bibitem[Chatterjee 
\& Tan(2014)]{cha14} Chatterjee, S., \& Tan, J.~C.\ 2014, \apj, 780, 53 

\bibitem[Chiang 
\& Laughlin(2013)]{chi13} Chiang, E., \& Laughlin, G.\ 2013, \mnras, 431, 3444 

\bibitem[Dong et al.(2014)]{don14} Dong, S., Zheng, Z., Zhu, 
Z., et al.\ 2014, \apjl, 789, L3 

\bibitem[Dressing 
\& Charbonneau(2013)]{dressingcharbonneau2013} Dressing, C.~D., \& Charbonneau, D.\ 2013, \apj, 767, 95 

\bibitem[Fabrycky et al.(2012)]{fabryckyetal2012} Fabrycky, D.~C., Ford, 
E.~B., Steffen, J.~H., et al.\ 2012, \apj, 750, 114 

\bibitem[Fischer 
\& Valenti(2005)]{fis05} Fischer, D.~A., \& Valenti, J.\ 2005, \apj, 622, 1102 

\bibitem[Foreman-Mackey et al.(2014)]{for14} Foreman-Mackey, 
D., Hogg, D.~W., \& Morton, T.~D.\ 2014, \apj, 795, 64 

\bibitem[Fressin et al.(2013)]{fre13} Fressin, F., Torres, 
G., Charbonneau, D., et al.\ 2013, \apj, 766, 81 

\bibitem[Hansen 
\& Murray(2013)]{han13} Hansen, B.~M.~S., \& Murray, N.\ 2013, \apj, 775, 53 

\bibitem[Johnson et al.(2010)]{joh10} Johnson, J.~A., Aller, 
K.~M., Howard, A.~W., \& Crepp, J.~R.\ 2010, \pasp, 122, 905 

\bibitem[Kasting et al.(1993)]{kastingetal1993} Kasting, J.~F., 
Whitmire, D.~P., \& Reynolds, R.~T.\ 1993, \icarus, 101, 108 

\bibitem[Kopparapu et al.(2014)]{kopparapuetal2014} Kopparapu, R.~K., 
Ramirez, R.~M., SchottelKotte, J., et al.\ 2014, \apjl, 787, L29

\bibitem[Kopparapu et al.(2013)]{kopparapuetal2013} Kopparapu, R.~K., 
Ramirez, R., Kasting, J.~F., et al.\ 2013, \apj, 765, 131 

\bibitem[Kane(2014)]{kane2014} Kane, S.~R.\ 2014, \apj, 782, 111 

\bibitem[Kane et al.(2012)]{kaneetal2012} Kane, S.~R., Ciardi, 
D.~R., Gelino, D.~M., \& von Braun, K.\ 2012, \mnras, 425, 757 

\bibitem[Kane \& Gelino(2012a)]{kanegelino2012} Kane, S.~R., 
\& Gelino, D.~M.\ 2012, \pasp, 124, 323

\bibitem[Kane 
\& Gelino(2012b)]{kanegelino2012-2} Kane, S.~R., \& Gelino, D.~M.\ 2012, Astrobiology, 12, 940 

\bibitem[Kane \& von Braun(2008)]{kanevonbraun2008} Kane, S.~R., 
\& von Braun, K.\ 2008, \apj, 689, 492 

\bibitem[Kane et al.(2013)]{kanehinkel2013} Kane, S.~R., Hinkel, 
N.~R., \& Raymond, S.~N.\ 2013, \aj, 146, 122 

\bibitem[Lissauer et al.(2011)]{lissaueretal2011} Lissauer, J.~J., 
Ragozzine, D., Fabrycky, D.~C., et al.\ 2011, \apjs, 197, 8 

\bibitem[Lissauer et al.(2012)]{lissaueretal2012} Lissauer, J.~J., 
Marcy, G.~W., Rowe, J.~F., et al.\ 2012, \apj, 750, 112

\bibitem[Lissauer et al.(2014)]{lis14} Lissauer, J.~J., 
Marcy, G.~W., Bryson, S.~T., et al.\ 2014, \apj, 784, 44 

\bibitem[Lopez \& Fortney(2014)]{lopezfortney2014} Lopez, E.~D., \& Fortney, J.~J.\ 2014, \apj, 792, 1 

\bibitem[Mordasini et al.(2012)]{mordasinietal2012} Mordasini, C., 
Alibert, Y., Georgy, C., et al.\ 2012, \aap, 547, A112 

\bibitem[Mullally et al.(2015)]{mul15} Mullally, F., 
Coughlin, J.~L., Thompson, S.~E., et al.\ 2015, \apjs, 217, 31 

\bibitem[Petigura et al.(2013)]{pet13} Petigura, E.~A., 
Howard, A.~W., 
\& Marcy, G.~W.\ 2013, Proceedings of the National Academy of Science, 110, 19273 

\bibitem[Quintana et al.(2014)]{quintanaetal2014} Quintana, E.~V., 
Barclay, T., Raymond, S.~N., et al.\ 2014, Science, 344, 277 

\bibitem[Rein(2012)]{rei12} Rein, H.\ 2012, \mnras, 427, L21 

\bibitem[Rowe et al.(2014)]{roweetal2014} Rowe, J.~F., Bryson, 
S.~T., Marcy, G.~W., et al.\ 2014, \apj, 784, 45 

\bibitem[Rowe et al.(2015)]{row15} Rowe, J.~F., Coughlin, 
J.~L., Antoci, V., et al.\ 2015, \apjs, 217, 16 

\bibitem[Santerne et 
al.(2012)]{san12} Santerne, A., D{\'{\i}}az, R.~F., Moutou, C., et al.\ 2012, \aap, 545, A76 

\bibitem[Sousa et 
al.(2011)]{sou11} Sousa, S.~G., Santos, N.~C., Israelian, G., Mayor, M., \& Udry, S.\ 2011, \aap, 533, A141 

\bibitem[Swift et al.(2013)]{swi13} Swift, J.~J., Johnson, 
J.~A., Morton, T.~D., et al.\ 2013, \apj, 764, 105 

\bibitem[Torres et al.(2011)]{tor11} Torres, G., Fressin, F., 
Batalha, N.~M., et al.\ 2011, \apj, 727, 24 

\bibitem[Williams 
\& Pollard(2002)]{wil02} Williams, D.~M., \& Pollard, D.\ 2002, International Journal of Astrobiology, 1, 61

\bibitem[Wright et al.(2011)]{wrightetal2011} Wright, J.~T., Fakhouri, 
O., Marcy, G.~W., et al.\ 2011, \pasp, 123, 412 
\end{thebibliography}
\end{document}